\documentclass{vienna-conf2019}
\usepackage{graphicx}
\usepackage{xcolor}
\usepackage{hyperref}
\usepackage[]{natbib}  
\usepackage{epstopdf}

\def\BibTeX{{\rm B\kern-.05em{\sc i\kern-.025em b}\kern-.08em
    T\kern-.1667em\lower.7ex\hbox{E}\kern-.125emX}}
\bibpunct{(}{)}{;}{a}{}{,}  
\newcommand{\kms}{{\mathrm{km~s^{-1}}}}
\newcommand{\ms}{{\mathrm{m~s^{-1}}}}

\newcommand{\gecervs}{\textit{GE-CeRVS}}

\begin{document}

\TitreGlobal{Stars and their variability observed from space}


\title{Cepheids under the magnifying glass $-$ not so simple, after all!}

\author{R.~I. Anderson}\address{European Southern Observatory, Karl-Schwarzschild-Str. 2, 85748 Garching b. M\"unchen, Germany}

\setcounter{page}{237}


\maketitle


\begin{abstract}
Classical Cepheids are blue loop stars that have famously been dubbed ``magnifying glasses of stellar evolution'' and have been studied for a long time. As more and more precise observations of Cepheids are secured over ever-increasing temporal baselines, our ignorance of the physics governing these crucial stars is becoming increasingly clear. Thus, it is time to turn up the magnification and investigate the limitations of our understanding of classical pulsators. 
Of course, classical Cepheids are also standard candles thanks to the Leavitt law that allows to measure distances in the nearby Universe. Nowadays, Cepheids serve as the backbone of a precisely calibrated distance ladder that allows to measure Hubble's constant ($H_0$) to better than $2\%$, thus providing crucial constraints for precision cosmology. The recently established discord among $H_0$ values measured in the late-time Universe and inferred from observations of the early Universe requires utmost diligence in estimating systematic uncertainties in order to strengthen the significance of the results. 
In this presentation I focus on two main aspects of recent Cepheid-related research. First, I present the Geneva Cepheid Radial Velocity survey (GE-CeRVS) and an update on the  modulated spectroscopic variability exhibited by the 4\,d Cepheid QZ~Normae based on 8 years of monitoring. Then, I discuss some efforts directed towards a 1\% $H_0$ measurement needed for understanding the cosmological implications of discordant $H_0$ values. Finally, I argue that now is a particularly opportune time to leverage the synergies between stellar physics and observational cosmology.
\end{abstract}



\section{Introduction}

Classical Cepheids (henceforth: Cepheids) are evolved intermediate-mass high-amplitude radially pulsating stars with important applications for stellar astrophysics and cosmology. The variability of these very luminous stars provides insights into a stellar evolutionary phase that is notoriously difficult to model \hbox{---} the blue loop during core He burning \hbox{---} and allows to determine precise (extragalactic) distances thanks to the Leavitt law \citep{Leavitt1912}. Importantly, Cepheids form the backbone of the currently most precise distance ladder used for measuring the local expansion rate of the Universe, $H_0$, which has become a focus of observational cosmologists, since recent precise measurements of $H_0$ differ by approximately $9\%$ depending on whether they are based on the late Universe (``today'') or the early Universe as it was $13.8$ billion years ago \citep[and cf. \citealt{Verde2019} for a recent conference summary]{Riess2019}.

Despite the long history of astronomical research involving Cepheids \citep{Goodricke1786} and the precision of the modern distance ladder, recent research has increasingly identified a wealth of phenomena that defy explanation and are relevant for many different sub-fields of astronomy \& astrophysics, including the evolution of multiple stars, the effects of rotation, convection, and other internal mixing processes, the relation between mass and luminosity, and the variability content of Cepheids, in particular concerning the regularity of Cepheid pulsations. Hence, I would argue that the adjective ``classical'' should not be considered synonymous with ``well-understood'', and that digging deeper to identify and understand these issues will help to progress to a better understanding of how stars and the cosmos evolve. Additionally, a reinforced astrophysical basis for the objects that calibrate the distance ladder is required for strengthening the interpretation of discordant $H_0$ values as an indication of new physics beyond the concordance model of cosmology, that is, flat $\Lambda$CDM.

\section{Modulated Spectroscopic Variability}
\subsection{The Geneva Cepheid Radial Velocity Survey}

The Geneva Cepheid Radial Velocity Survey (\gecervs, Anderson et al. in prep.) is a large ongoing survey dedicated to measuring high-precision radial velocities of Galactic classical Cepheids. Since 2011, \gecervs\ has gathered $> 19000$ observations of $> 300$ candidate Cepheids. \gecervs\ operates on two 1m-class telescopes that provide access to the full sky: the Flemish Mercator telescope located at the Roque de los Muchachos Observatory on the island of La Palma, Spain, and the Swiss Euler telescope at the La Silla site of the European Southern Observatory in Chile. Both telescopes feature efficient fiber-fed high-resolution optical echelle spectrographs \citep{Queloz2001,Raskin2011}. While RVs measured with Hermes reach approximately $10-15\,\ms$ precision, the simultaneous wavelength reference provided by Coralie's Fabry-P\'erot etalon affords a precision of $1-2\,\ms$ for bright objects. Thanks to their high precision, dense pulsation phase coverage, and multi-year baseline, \gecervs\ data provide an unprecedented view of the spectroscopic variability of Cepheids. Additionally, they provide a crucial reference for understanding RV variability measured by the RVS instrument on-board the \textit{Gaia} spacecraft \citep{Wallerstein2019}. Among other things, \gecervs\ data have shown that Cepheid RV signals are modulated on different timescales. Specifically, \citet{Anderson2014rv} showed that short-period (overtone?) Cepheids QZ~Nor and V335~Pup exhibit long-term changes in their RV amplitudes, whereas long-period Cepheids $\ell$~Car and RS~Pup exhibit cycle-to-cycle variations in pulsation period, RV amplitude, and line profiles \citep[cf. also][]{Anderson2016c2c}. 

The following subsection provides a work-in-progress update on this initial discovery 5 years ago and illustrates this surprising long-term behavior for the short-period Cepheid QZ~Nor. In addition to RV measurements, it considers the Bisector Inverse Span (BIS) \citep[for a visualization, see e.g.][their Fig.\,2]{Anderson2016c2c}, which provides a useful and precise measure of spectral line asymmetry.

\subsection{Monitoring QZ~Nor for 8 years}

\begin{figure}
\centering
\includegraphics[width=1\textwidth]{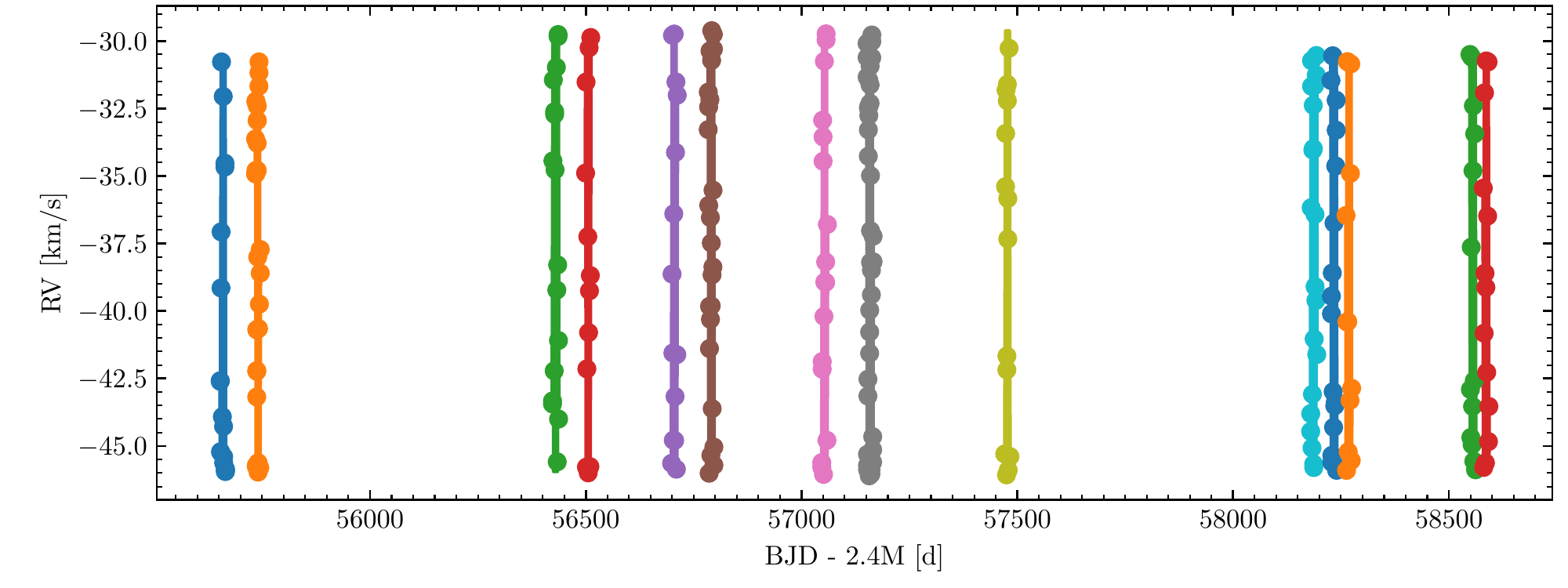}
\includegraphics[width=1\textwidth]{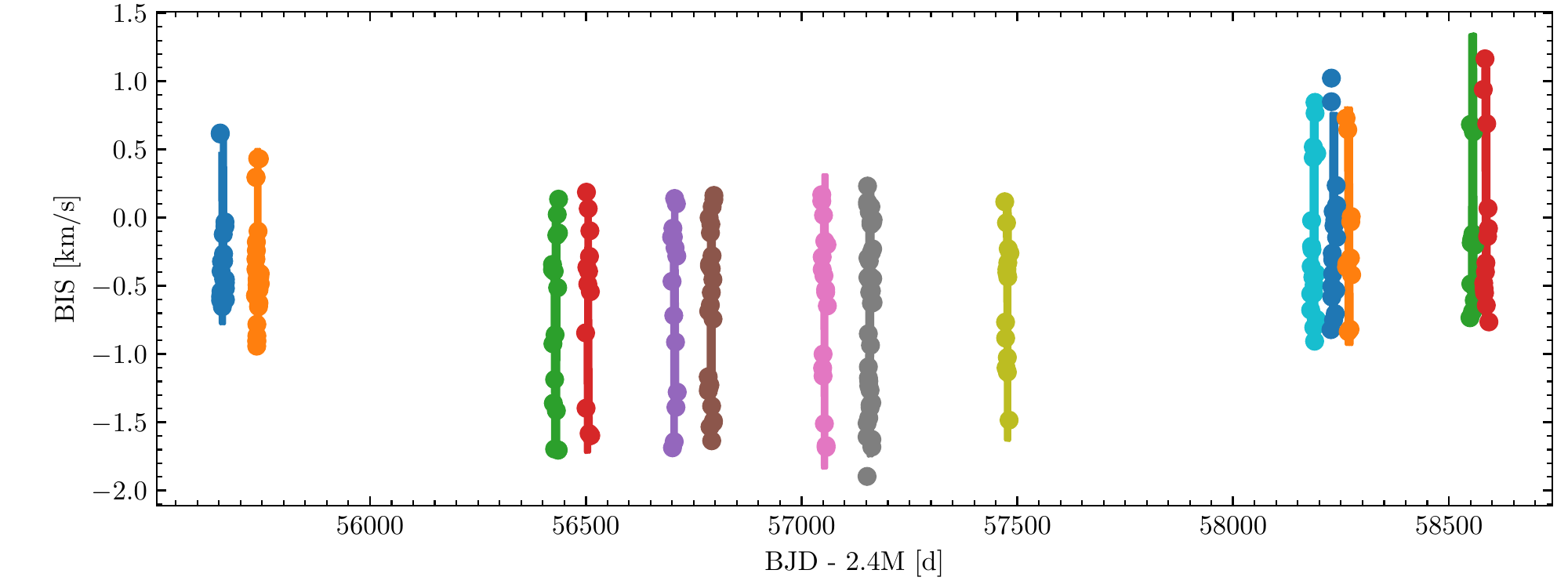}
\caption{GE-CeRVS data of the $3.79$\,d (overtone?) Cepheid QZ~Nor. \textbf{top:} RV against observation date. \textbf{bottom:} Line asymmetry as measured by BIS against observation date
\label{fig:QZNorTimeSeries}}
\end{figure}

Figure\,\ref{fig:QZNorTimeSeries} shows the \gecervs\ time series RV and BIS measurements of the $3.79$\,d Cepheid QZ~Nor, whose modulated RV curve was first reported by \citet{Anderson2014rv}. As is clear from the figure, both the RV amplitude and the line profile asymmetry traced by the BIS quantity exhibit long-term changes throughout the 8-year time scale of the observations.

\begin{figure}
\begin{minipage}{0.5\textwidth}
\includegraphics[width=1\textwidth]{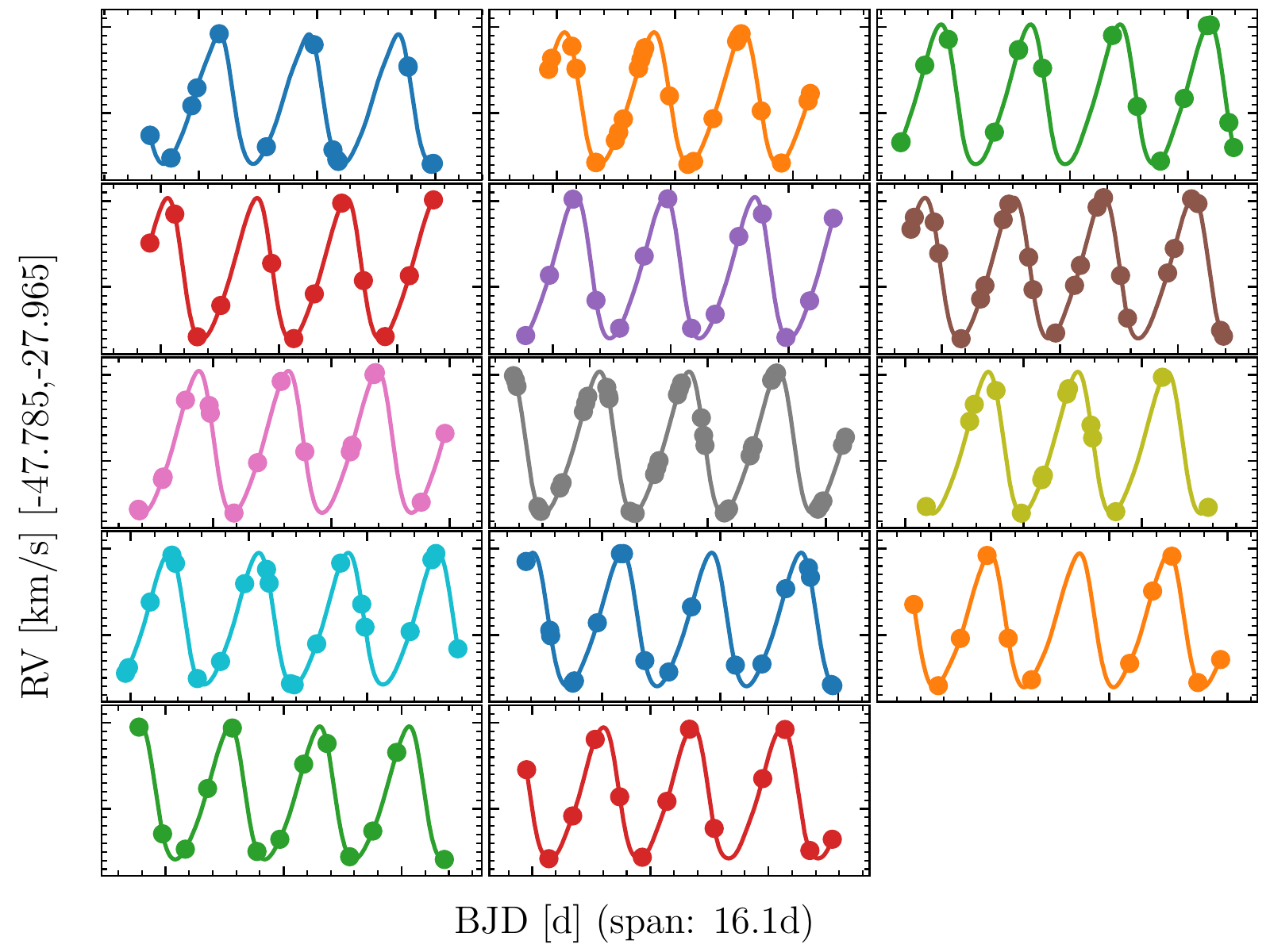}
\caption{QZ Nor's RV curves per epoch
\label{fig:QZNorPerEpoch}}
\end{minipage}
\begin{minipage}{0.5\textwidth}
\includegraphics[width=1\textwidth]{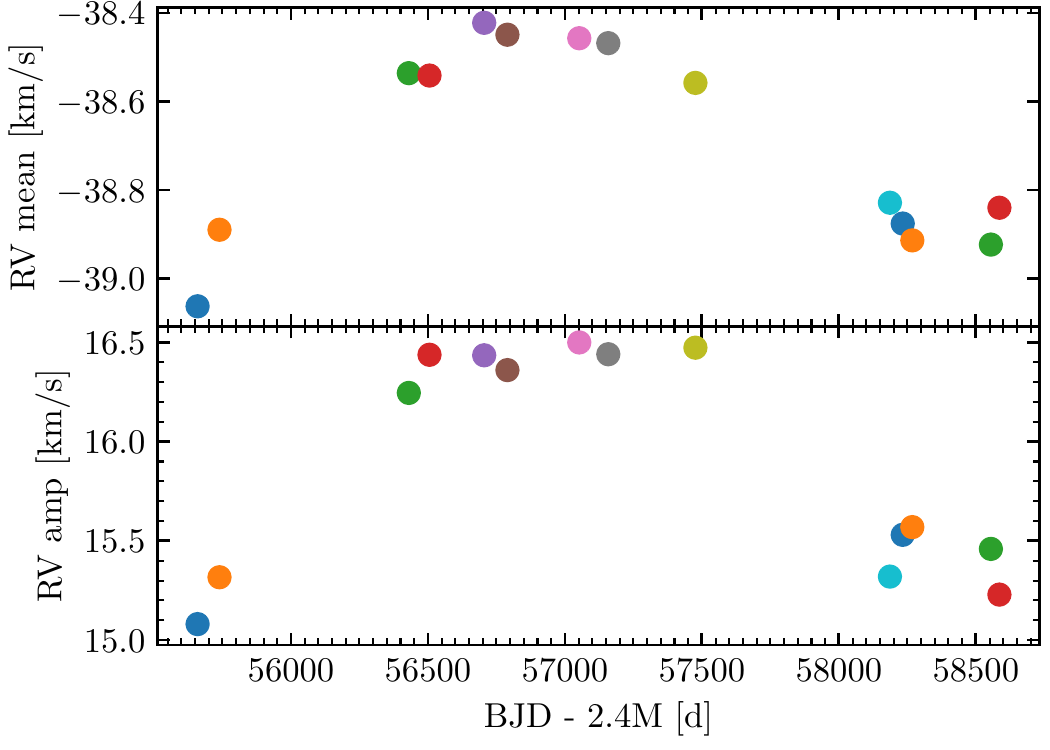}
\caption{QZ Nor's mean RV and RV amplitude variations over time
\label{fig:QZNorMeanAmp}}
\end{minipage}
\end{figure}

Figure\,\ref{fig:QZNorPerEpoch} separates the RV variability into epochs of $16.1$\,d duration and illustrates that each epoch is reasonably well sampled, exhibiting fairly simple variability curves that can be represented by low-order Fourier series. However, Fig.\,\ref{fig:QZNorMeanAmp} shows that the pulsation-averaged RV ($v_\gamma$) and the RV amplitude ($A_{\mathrm{RV}}$) exhibit correlated changes over time. This is important because the top panel in Fig.\,\ref{fig:QZNorMeanAmp} resembles a low-amplitude high-eccentricity orbit, and would likely be interpreted as such if the bottom panel did not imply a contemporaneous change in $A_{\mathrm{RV}}$.  
In other words, Fig.\,\ref{fig:QZNorMeanAmp} cautions us against interpreting all temporal variations on the order of even $1\,\kms$ as signs of orbital motion, contradicting the typical interpretation of such signals. At present, it remains unclear whether these changes are (quasi-)periodic or repeating, not to mention what causes these phenomena. \citep{Derekas2017} reported the previously longest modulation seen in a Cepheid for the only classical Cepheid in the original \textit{Kepler} field, V1154~Cyg, with a period of $1160$\,d  based on a light curve spanning $1470$\,d. Unfortunately, we do not yet know whether photometric and spectroscopic modulations correspond to each other. This is mostly because precise long-timescale photometric monitoring of bright stars is not available, whereas spectroscopic observations do not exist in sufficient quantity for fainter Cepheids, e.g. in the OGLE fields \citep[e.g.][]{Soszynski2019}. However, \citet{Anderson2016vlti} showed that interferometric observations suggest that $\ell$~Car exhibits cycle-to-cycle changes in its maximum diameter. High-quality contemporaneous photometric and spectroscopic time series are required for understanding these phenomena and casting open these exciting windows into stellar pulsations.

As explained in \citet{Anderson2018rv}, different ways of measuring RV in Cepheids have different merits and potential issues. In \gecervs, RV is defined as the center of a Gaussian line fitted to the cross-correlation profile, which is the standard in precision RV measurements of stable stars \citep{Baranne1996,Pepe2002}. Astrophysical effects can change the shape of line profiles over time, out of sync with the dominant radial mode pulsation. Such effects can lead to spurious changes in the mean velocity by introducing time-dependence in the bias of RVs measured using Gaussian fits to asymmetric line profiles. Although this complicates the interpretation of $v_\gamma$ as an indicator for spectroscopic binary motion, this is useful for identifying previously unknown hydrodynamical effects, such as coupling between pulsations and convection \citep{Anderson2016c2c}. Another particularly puzzling, and potentially telling, signal is the $40.2$\,d periodic BIS variation seen in the North Star, Polaris \citep{Anderson2019polaris}, whose light curve and RV variations are dominated by periodic variability on a timescale of $3.97$\,d. 

\section{Towards a 1\% measurement of Hubble's Constant}

Very large efforts are under way to measure the local expansion rate of the Universe, Hubble's constant $H_0$ to within 1\%. Until recently, a prime motivation for pursuing this factor of 10 improvement over the Hubble Space Telescope ({\it HST}) Key Project \citep{Freedman2001} was an improved ability to elucidate the nature of dark energy. Indeed, a $1\%$ $H_0$ measurement used as prior for determining the equation of state of dark energy would optimally support future developments in observational cosmology \citet{Weinberg2013}. In as much as it is acceptable to use $H_0$ as a prior for the analysis of the Cosmic Microwave Background (CMB), this motivation remains true and important. However, recent works have found a conspicuous disagreement between  measurements of $H_0$ based on the late-time Universe (as it is ``today'' by cosmic standards) and the value of $H_0$ inferred from observations of early Universe physics, such as the CMB or Big Bang Nucleosynthesis \citep[cf.][for an overview]{Verde2019}. 
Notably, there is a $\sim 9\%$ difference (at $4.4\sigma$ significance) between a faster late-Universe $H_0$ and a slower early-Universe value based on an extragalactic distance scale composed of classical Cepheids and type-Ia supernovae \citep{Riess2019} and observations of the CMB by the {\it Planck} satellite \citep{Planck2018H0}. This \emph{Hubble tension} challenges the adequacy of the concordance cosmological model, which connects the present-day Universe with the oldest observed radiation and describes the Universe as flat, consisting of dark energy in form of a cosmological constant, dark matter, and ordinary matter ($\Lambda$CDM). If it can be shown that new physics are required to solve this conundrum, then we may well be near a breakthrough in our understanding of the Universe and its evolution.

Classical Cepheids play a crucial role for increasing the precision of $H_0$ measurements for several reasons. As easily recognizable and very luminous stars, Cepheids provide the absolute calibration of type-Ia supernovae, whose distance-redshift relation defines the Hubble-Lema\^itre law. Although Cepheid alternatives such as Mira stars \citep{Huang2019} and stars near the Tip of the Red Giant Branch \citep[TRGB]{Freedman2019} have made tremendous progress over the last few years, they have not yet reached the same precision as that currently offered by Cepheids and it remains to be seen which type of primary standard candle will first be limited by systematic uncertainties. Hence, it is crucial to determine the systematic ``error floor'' of Cepheids (and other standard candles) and to pay particular attention to previously neglected or overlooked systematic effects.

\citet{AndersonRiess2018} recently examined a frequently mentioned, albeit poorly quantified, systematic uncertainty of the distance ladder: stellar association bias. In particular, we considered the effect of stellar multiplicity and cluster membership on the modern $H_0$ measurements as implemented by \citet{Riess2016}, rather than describing the effects on the Leavitt law (or period-luminosity relation). The latter exercise provides a valuable test of stellar physics, that is whether stellar models correctly predict observed Leavitt laws. In terms of a systematic error or bias of $H_0$, however, it is most important to consider the equivalence between Leavitt laws observed in nearby and more distant Cepheid populations. 

\begin{figure}
\centering
\includegraphics{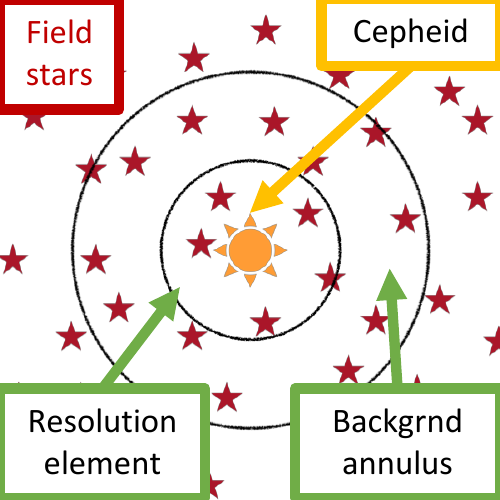}
\includegraphics{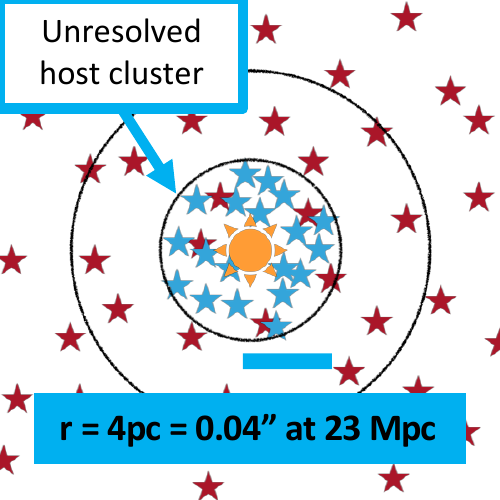}
\caption{Distinction between chance blending (left) and blending of physically associated stars that leads to biased extragalactic measurements and $H_0$ (right). Chance blending can be corrected statistically by measuring field star contributions. However, blending of physically associated stars cannot be corrected in this way, since the physical scale of stellar association in clusters is unresolved at typical supernova-host galaxy distances\label{fig:blending}.}
\end{figure}

The bias arises because Cepheids are relatively young (30-300 Myr old) evolved supergiant stars that occasionally in open star clusters or loose associations \citep[e.g.][]{Turner2010,Anderson2013}. On Local Group scales, star clusters are easily recognized and spatially resolved. At the average supernova host galaxy distance of 23 Mpc, however, the typical physical cluster scale of 4 pc is equal to the plate scale of {\it HST}'s WFC3 in UVIS mode (0.04"). Thus, stars physically associated with Cepheids do not contribute to the calibration of the Leavitt law, which is done locally, but do contribute light to distant Cepheids on the second rung of the distance ladder. Figure\,\ref{fig:blending} illustrates this. Stellar association bias is different from chance blending of field stars because the latter is corrected using artificial star tests, whereas cluster light contributions cannot be easily corrected in distant galaxies.

Using M31 as a supernova-host analog for its high metallicity, spiral structure, and external view, \citet{AndersonRiess2018} measured the light contributions for 9 cluster Cepheids using {\it  HST} photometry from the PHAT project \citep{Dalcanton2012}. Cluster positions were known from the Andromeda Project \citep{Johnson2015} and Cepheid positions from the PanSTARRS survey \citep{Kodric2018}, M31's cluster Cepheids were highlighted by \citet{Senchyna2015}. 

Interested readers are referred to \citet{AndersonRiess2018} for the details of the analysis. Suffice it here to state that the bias was estimated as the product of the clustered Cepheid fraction times the average bias measured using curve-of-growth analysis of 9 M31 cluster Cepheids. We find that on average a Cepheid's distance is underestimated by $7.4$ mmag using a reddening-free Wesenheit magnitude that combines $V$, $I$, and $H$-band photometry, and by $9.8$ mmag when using $H$-band only. Corrections for these light contributions are now included in the $H_0$ measurements by \citet{Riess2019}. Further work in this direction will clarify the adequacy of the adopted clustered Cepheid fraction and the effects of selection criteria applied to extragalactic Cepheid samples on this bias. 

Of course, Cepheids very frequently occur in multiple star systems \citep[e.g.][]{Evans2015rv,Kervella2019a,Kervella2019b}. However, companion stars contribute a bias of $< 0.004\%$ because most configurations of companion stars are rarely spatially resolved \citep{AndersonRiess2018}. This estimate was based on the Geneva stellar evolution models \citep{Ekstroem2012,Georgy2013,Anderson2016rot} and the properties of Cepheid orbits \citep{Moe2017}.

Another bias affecting the distance scale due to inherent differences between near and far Cepheid populations concerns the dilation of variability periods due to cosmological redshift. As shown in \citet{Anderson2019rlb}, time dilation has previously led to a $0.3\%$ underestimate of $H_0$. However, this bias will increase as future work will focus on more and more distant Cepheids in order to increase the number of supernova-host galaxies, which is limited by the rate of supernova explosions in the nearby Universe. For reference, Leavitt law distances to pulsating stars at $100$\,Mpc would be biased by $2\%$ if redshift is not accounted for. Thus, future distance scale work aimed at a $1\%$ measurement of $H_0$ must account for dilated variability periods, for example using observed host galaxy redshifts.

\section{Synergies between precision cosmology and stellar physics}

Extragalactic pulsating stars (e.g. Cepheids \& Miras) provide the backbone of the extragalactic distance ladder and the associated measurement of $H_0$ sets the scale and age of the Universe. As we are pursuing a $1\%$ $H_0$ measurement (or better), additional scrutiny is required to ensure that systematic uncertainties that could shift the center value of $H_0$ are under control at the same level, or better. Given that the primary distance ladder rungs dominate the uncertainty budget of the $H_0$ measurement \citep[cf.][]{Riess2019}, further scrutiny is required to identify and mitigate systematics affecting Cepheid distances. This leads to wonderful opportunities for studying populations of extragalactic Cepheids in different environments (star formation, galactic potentials, chemical composition, etc.) that can provide new insights into stellar pulsations and the evolution of stellar populations.

At the same time, precision observations of the closest Cepheids reveal surprising new phenomena, such as multi-periodic variability (cf. above) or difficult-to-explain circumstellar environments \citep{Hocde2019}. Moreover, long-term photometric monitoring of Cepheid pulsation periods challenges the canonical interpretation of observed rates of period change as indicative of secular evolution \citep[e.g.][]{Poleski2008,Sueveges2018a}. These are fundamental new insights into stellar astrophysics, which ultimately explain the mechanisms upon which the distance ladder rests. Of course, mmag-level non-radial pulsations and short-term period fluctuations do not immediately lead to biases in measuring $H_0$. However, such work may allow us to select cleaner samples of Cepheids exhibiting tighter Leavitt laws that ultimately increase distance ladder precision and accuracy. With large time-domain surveys such as {\it Gaia} and LSST (cf. contributions by L. Eyer and G. Clementini in these proceedings) gathering unprecedented data for classical pulsators, and even small telescopes providing high-quality datasets (such as GE-CeRVS), now is a great time to leverage the synergies between precision cosmology and detailed stellar physics.

\begin{acknowledgements}
I would like to thank the organizers, and Werner Weiss in particular, for this wonderful and inspiring meeting and my collaborators, especially everyone involved with GE-CeRVS, incl. L. Eyer and N. Mowlavi.
\end{acknowledgements}

\bibliographystyle{aa}  
\bibliography{Anderson_3k07} 

\end{document}